\documentclass[twocolumn,nofootinbib,showpacs,showkeys,prd]{revtex4}
\usepackage{amsmath}
\usepackage{amsfonts}
\usepackage{amssymb}
\usepackage{graphicx}
\usepackage{color}

\begin{document}

\title{Dark energy from entanglement entropy}

\author{Salvatore Capozziello}
\email{capozzie@na.infn.it}
\affiliation{Dip. di Fisica, Universit\`a di Napoli "Federico II",
             Via Cinthia, I-80126, Napoli, Italy}
\affiliation{INFN Sez. di Napoli, Compl. Univ. Monte S. Angelo Ed. N
Via Cinthia, I- 80126 Napoli, Italy.}

\author{Orlando Luongo}
\email{orlando.luongo@na.infn.it}
\affiliation{Instituto de Ciencias Nucleares, Universidad Nacional Aut\'onoma de M\'exico, AP 70543,
M\'exico DF 04510, Mexico;}
\affiliation{Dip. di Fisica, Universit\`a di Napoli "Federico II",
             Via Cinthia, I-80126, Napoli, Italy}
\affiliation{INFN Sez. di Napoli, Compl. Univ. Monte S. Angelo Ed. N
Via Cinthia, I- 80126, Napoli, Italy.}

\begin{abstract}
We show that  quantum decoherence, in the context of observational cosmology, can be connected to   the cosmic  dark energy. The decoherence signature could be characterized by the existence of quantum entanglement between cosmological eras. As a consequence, the Von Neumann entropy related to  the entanglement process, can be compared to the thermodynamical entropy in a homogeneous and isotropic universe. The corresponding cosmological models are compatible with the current observational bounds being able to reproduce viable equations of state without introducing {\it a priori}  any cosmological constant. In doing so, we investigate two cases, corresponding to two suitable cosmic volumes, $V\propto a^3$ and $V\propto H^{-3}$,  and  find  two models which fairly well approximate the current cosmic speed up. The existence of dark energy can be therefore reinterpreted as a quantum signature of entanglement, showing that the  cosmological constant represents a limiting case  of a more complicated model derived from the quantum decoherence.
\end{abstract}


\maketitle

\section{Introduction}
\label{sec:int}

The issue of unifying both classical and quantum cosmology is an open challenge of modern physics partially due to the lack of a  self-consistent quantum gravity theory. All the attempts aimed to  achieve a quantum gravity theoretical scheme lead to many inconsistencies and it is furthermore not clear how to measure  quantum signatures in the context of observational cosmology \cite{klaus}. Recently, the evidence of  cosmic positive acceleration,  reported by several observational surveys \cite{sn0,sn1,sn2}, has open further interrogatives, due to the ignorance of the physical nature of the fluid responsible for the cosmic speed up. The standard framework leads to the existence of a vacuum energy cosmological constant $\Lambda$, characterized by a negative equation of state \cite{coppa,coppa2}. Even though  the cosmological constant allows the cosmic acceleration  at late times, the observational bounds on $\Lambda$ are incompatible with theoretical predictions of a gravitational vacuum state. In addition, the $\Lambda$ and matter order of magnitude are surprisingly close to each other, although $\Lambda$ is supposed to be constant along the universe evolution. These two thorny shortcomings, namely the {\it fine tuning} and the {\it coincidence} problems, disturb the otherwise appealing picture of a cosmological constant and dramatically plague  the so called  $\Lambda$CDM model \cite{coppa3,coppa4}.

A way out for the above problems  is to extend the standard  $\Lambda$ paradigm by postulating the existence of a new ingredient referred to as {\it dark energy}. It provides a fluid with an evolving negative equation of state. A  solution is that quantum effects at fundamental level could be responsible for the existence of dark energy \cite{ciao}. The possibility to relate dark energy with quantum effects has been   recently considered by dealing with the existence of a hidden mechanism behind the nature of the cosmological constant as a result of entanglement between cosmological eras \cite{io}. In doing so, the existence of entangled states was postulated, showing that a running cosmological constant is compatible with early times of the universe evolution, while, at late times, the cosmological constant is recovered as a limiting case. Such a result can be achieved also in the framework of alternative theories of gravity \cite{rept}. A key role for this approach is played by  quantum information that reached a growing   interest during  last decades spanning from condensed matter to cosmology \cite{udso1,udso2,udso3,udso4,udso5,udso6,udso7}. Moreover, relevant  entanglement phenomena have been framed into robust theoretical schemes and  verified through several experiments \cite{exp}.
Thus,  selection criteria were sought  in order to characterize the entanglement amount of  quantum states. All of those criteria are essentially based on equivalent forms of non-locality of pure quantum states. In addition, since entanglement can be considered as a  fundamental quantity, its physical meaning has been compared with  quantities like energy and entropy.

The purpose of this work is twofold: we first wonder whether it is possible to consider cosmic  dark energy as the result of an entanglement mechanism between two or more than two cosmological eras during the universe evolution. Secondly, we  show that the corresponding equation of states, derived by  the entanglement process, is compatible with the late time universe evolution as a running barotropic fluid equation of state. This approach is  justified assuming  the entanglement mechanism applicable  to the  degrees of freedom  of macroscopic systems, in this case, the whole observed universe. The latter process is known in the literature as decoherence \cite{sett}. In particular, the challenge of finding decoherence effects in the cosmological scenario is related to the interaction between qubits with their environment \cite{sett2}. In particular,  quantum coherence and entanglement of the quantum states can lead to {\it more decoherence} if the number of qubits increases \cite{sett3}.

In this paper, under the hypothesis that the decoherence effects could be related to dark energy, we show that it is possible to associate the Von Neumann entropy, as a naive selection criterium, to the fluid accelerating the universe. Choosing the simplest case in which the dark energy represents the measurement of decoherence in the observable universe, we demonstrate that it is possible to infer a cosmological model which predicts the dark energy effects without introducing a vacuum energy cosmological constant term at late times. The paper is structured as follows: in Sec. II we discuss the thermodynamics of a homogeneous and isotropic universe in the context of entanglement. In Sec. III we describe the cosmological models derived by considering the entanglement decoherence as a source for dark energy. In Sec. IV, we draw the conclusions and discuss the perspectives of the present approach.

\section{Thermodynamics and entanglement in cosmology}

Let us summarize the main features of entanglement between cosmological eras from the thermodynamic point of view. We pay particular attention to the relation of   entanglement effects with  thermodynamics of the observable universe. In doing so, it is prominent to start by taking into account  an $N$-dimensional Hilbert space whose probability of finding an observable is $p_k = \frac{1}{N}$, $\forall k$. The particular case in which only one of the $p_k$ is different from zero corresponds to a pure quantum state. We are interested in
non-local and entangled states; therefore, hereafter we assume  that $p_k\neq 0$ for a given set of $k$.
We introduce the density matrix $\hat\rho$ associated to the quantum states. For a generic pure state, we have
\begin{equation}\label{matriosca}
\hat\rho_p=|\Psi\rangle\langle\Psi|\,.
\end{equation}
with the conditions $Tr\hat\rho=1$, $Tr\hat\rho^2\leq1$, $\langle\chi|\hat\rho|\chi\rangle\geq 0$, and $Tr\hat\rho^2=1$ if the state is
pure, $Tr\hat\rho^2<1$ otherwise \cite{ancora}.
In the case of non-pure state, the generalization of Eq. ($\ref{matriosca}$) is
\begin{equation}\label{genmatriosca}
\hat\rho=\sum_j|\Psi_j\rangle\langle\Psi_j|\,.
\end{equation}
We are interested in finding selection criteria able to quantify the entanglement degree in a given quantum state. As a first example, we can consider the basic concept of \emph{linear} \emph{entropy} $S_L=\frac{N}{N-1}\left(1-\mu\left[\hat\rho\right]\right)$. By interpreting $\mu[\rho]$ as a Taylor first order term of  $\hat\rho$, we can consider, as a widely used estimator of quantum correlations of a given subsystem, the so called Von Neumann entropy, defined as
\begin{equation}
\label{enggg}
S_{VN}=-Tr\left(\hat{\rho}\ln\hat{\rho}\right)\,=\,-\sum_k \lambda_k\ln
\lambda_k\,.
\end{equation}
In the framework of the Friedmann-Robertson-Walker (FRW) spatially flat metric, i.e.
\begin{equation}\label{hgjhg}
ds^2=dt^2-a(t)^2\left(dr^2+r^2\sin^{2}\theta d\phi^2+r^2d\theta^2\right)\,,
\end{equation}
one can model Eq. ($\ref{enggg}$) for different cosmological eras. Assuming that the eigenvalue $\lambda_K$, with $K\in k$ a fixed index, is the entropy dominating term, one can recast Eq. ($\ref{enggg}$) to be
\begin{equation}\label{nyafa}
S_{VN}=\rho\ln\rho\,,
\end{equation}
where $\rho$ represents an effective density coming from  the superposition of quantum sates. We motivate the idea of passing from Eq. ($\ref{enggg}$) to Eq. ($\ref{nyafa}$), by assuming dark energy as a lacking of information of quantum cosmological states. Hence, by following \cite{io}, $\rho$ can be interpreted as a dark energy density. Thus, the idea of considering entanglement, as a source of dark energy, could be justify as a quantum signature of cosmological states. In particular, the universe dynamics can be reproduced through entangled quantum states, in which current observable evidences are inferred from quantum decoherence.

In   cosmology, the dark energy term as well as the standard matter term are modeled by a perfect fluid energy-momentum tensor of the form
\begin{equation}
 T^{\alpha \beta}=(\rho +P)u^{\alpha} u^{\beta} - P g^{\alpha \beta}\,,
\end{equation}
where $g_{\alpha\beta}$ is the FRW metric, $u^\alpha$ the four velocity and $P\equiv\omega\rho$ the equation of state relating the pressure $P$ and the density $\rho$. In the equation of state for dark energy, $\omega$  is a negative number (function) that  guarantees the repulsive  effect counterbalancing the attraction of gravity. Considering the conservation law given by the contracted Bianchi identities $\nabla^\alpha T_{\alpha\beta}=0$, one gets the relation
\begin{equation}\label{utyyyy}
\frac{d\rho}{dz}=3\left(\frac{\omega+1}{1+z}\right)\rho\,,
\end{equation}
where  the scale factor of the universe has been reported in terms of the redshift  $a=1/(1+z)$. From the Friedmann equations,
\begin{eqnarray}\label{ave2}
H^2 &=& {8\pi G\over3}\rho\,,\nonumber\\
\,\\
\dot H + H^2&=&-{4\pi
G\over3}\left(\rho+3P\right)\,,\nonumber
\end{eqnarray}
where $H=\dot a/a$ is the Hubble parameter, and the first principle of thermodynamics \cite{calore}, we get the functional forms
\begin{equation}\label{ro}
 \rho \propto \exp\left[{3\int{\frac{1 + \omega(z)}{1+z}}dz}\right]\,.
\end{equation}
and
\begin{equation} \label{temp}
T \propto  \exp\left[{3\int{\frac{ \omega(z)}{1+z}}dz}\right]\,,
\end{equation}
for the density and temperature respectively in terms of $\omega$ as a function of the redshift $z$. Even though the dark energy and matter temperatures, in principle, could respectively dominate some epochs of the universe, no significative  modifications are today expected. We will make use of Eqs. ($\ref{ro}$) and ($\ref{temp}$) in order to find out an expression for $\omega$ in the case in which the dark energy term is derived from decoherence.

Here we assume that the Von Neumann entropy, as a measure of entanglement, contributes the whole entropy of the universe; in this context, it is possible to address the question of relating dark energy to decoherence.

\section{Dark energy from entanglement}

In this section, we discuss the hypothesis to relate the dark energy to the decoherence in the observable universe. In particular, as shown in \cite{io}, it is possible to define the  Von Neumann entropy in terms of observable cosmological quantities; thus, considering  that the eingenvalues of the Von Nuemann entropy can be derived  from the first principle of thermodynamics, we assume

\begin{equation}\label{kjefw}
\frac{dS_{VN}}{dz}\Big|_{z=0}=\frac{dS}{dz}\Big|_{z=0}\,,
\end{equation}
where $S$ is the thermodynamical entropy in terms of redshift. This position can be considered  the signature of decoherence at our epoch ($z=0$). Eq. ($\ref{kjefw}$) suggests that   dark energy  evolves as entropy and the whole entropy contribution  is equivalent to the one predicted by entanglement between states. Extending this postulate to redshift different from zero, i.e.

\begin{equation}\label{kjefwcoss}
\frac{dS_{VN}}{dz}\Big|_{z>0}=\frac{dS}{dz}\Big|_{z>0}\,,
\end{equation}
and by solving Eq. ($\ref{kjefw}$), considering from Eq. ($\ref{nyafa}$) the Von Neumann entropy first derivative with respect to $z$, we easily get
\begin{equation}\label{rhox}
-\frac{d\rho}{dz}\left(\log\rho+1\right)=\frac{1}{T}\frac{d\Big[\left(P+\rho\right)V\Big]}{dz}\,,
\end{equation}
which is the  continuity equation for a given volume $V$. Since such an equation contains the equation of state in its right hand side,  the dark energy behavior can be   derived from entanglement.
In order to solve Eq. ($\ref{rhox}$), we consider that the cosmological volume is defined in a standard thermodynamical way  by assuming $V=V_0R^{3}$, with $R$  a suitable  {\it universe radius}.  $R$ can be characterized in two different way. The first, as discussed in  \cite{calore}, is $R\propto a(t)$, leading to a volume which becomes zero at high redshift and $V_0\equiv\frac{1}{H_0^3}$ when $z=0$. For the second case, we assume that the universe size behaves as a comoving radius in the Hubble sphere. This assumption has been  investigated in
\cite{entr} and was firstly developed in the case of the Entropic Principle, suggesting that the radius of the universe scales as the inverse of the Hubble rate, that is $R\propto\frac{1}{H}$ \cite{entr2}. Assuming  $V^{(1)}\propto a^{3}$ and  $V^{(2)}\propto H^{-3}$, we get  from Eq.(\ref{rhox}), respectively
\begin{eqnarray}\label{tqjois}
3\frac{\log(e\rho)}{a^{3}}+\frac{1}{T}\Big[3 \omega(a)+\frac{1}{a}\frac{d\log(1+\omega(a))}{dz}\Big]=0\,,
\end{eqnarray}
and
\begin{eqnarray}\label{tqjois2}
3\frac{\rho}{a}\log(e\rho)(1+\omega(a))-\frac{1}{a\rho^2 T}\Big[6 (1+ \omega(a))^2-\frac{1}{a}\frac{d\omega}{dz}\Big]=0\,,\,\,\,
\end{eqnarray}
where $e$ is the Napier number. Eqs. ($\ref{tqjois}$) and ($\ref{tqjois2}$) have no analytic  solutions; by integrating numerically $\omega(a)$ in terms of the redshift $z$, we obtain  results represented  in Fig. (1) for  $\omega^{(1)}$ and $\omega^{(2)}$.

\begin{widetext}
\begin{center}
\begin{figure}[ht]
\includegraphics*[scale=1.4]{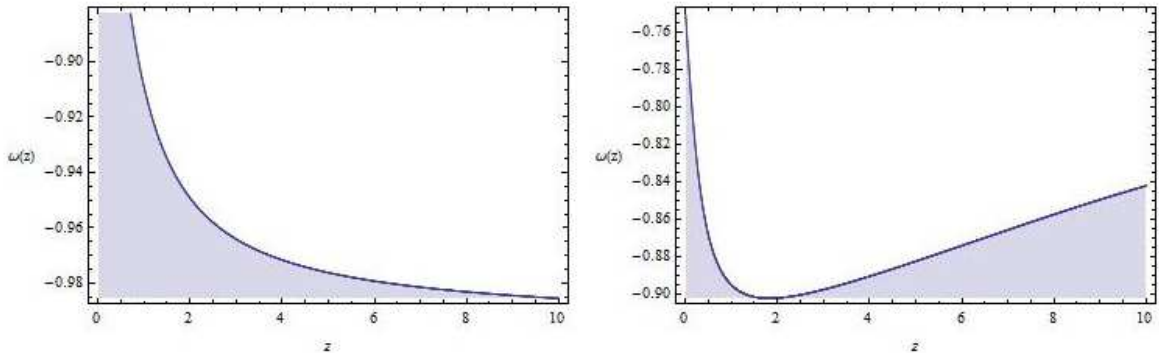}
\caption{Numerical plots of $\omega$ for the two cases of Eqs. ($\ref{tqjois}$) and ($\ref{tqjois2}$);  the case  $V=V_0a^{3}$  is plotted in  the left panel  and the case $V=V_0H^{-3}$ in  the right panel. Numerical coefficients for $\rho$, $T$ and $V$ are normalized to 1.}\label{3q1}
\end{figure}
\end{center}
\end{widetext}

The limiting behaviors for the  two equations of states in  Eqs. ($\ref{tqjois}$) and ($\ref{tqjois2}$) are  respectively:
\begin{eqnarray}
  \omega_{0}^{(1)} &=& -0.75\,, \nonumber\\
  \,\\
  \omega_{\infty}^{(1)} &\rightarrow& -1\,,\nonumber
\end{eqnarray}
and
\begin{eqnarray}
  \omega_{0}^{(2)} &=& -0.75\,, \nonumber\\
  \,\\
  \omega_{\infty}^{(2)} &\rightarrow& -0.85\,.\nonumber
\end{eqnarray}
It is worth noticing that, at high redshift,  that is for $z\gg1$, we recover the cosmological constant for $\omega^{(1)}$ as a limiting case. On the other hand, we infer that a dark energy behavior is consistent also for $\omega^{(2)}$. However, in  both  models, a pure cosmological constant is disfavored at low redshift; thus, it is clear that Eqs. ($\ref{tqjois}$) and ($\ref{tqjois2}$) provided a dark energy term, different from quintessence and from $\Lambda$ strictly holding at all epochs. In other words, while the cosmological constant could be seen as a limiting case at higher redshift regime for the first case, it is not recovered in the second case at any redshift.

The functional form of $\omega$ can be approximated by the following functions

\begin{equation}\label{w1}
\omega^{(1)}\sim \alpha + \beta \frac{1}{\beta + \gamma(1+z)^n}\,,
\end{equation}
and
\begin{equation}\label{w2}
\left\{
  \begin{array}{ll}
    \omega^{(2)}\sim -\delta^2 z, & \hbox{$0 \leq z\leq 1$;} \\
    \omega^{(2)}\sim \epsilon^2 z, & \hbox{$z\geq 1$.}
  \end{array}
\right.
\end{equation}
respectively, where $\alpha, \beta, \gamma, \delta, \epsilon$ are constants to be determined by observations.

Looking at Eqs. ($\ref{w1}$) and ($\ref{w2}$),  it is clear that by setting the initial conditions at $\omega(z=0)=-1$, the cosmological constant dominates over all the  eras of the universe evolution; on the contrary, the functional forms that approximate the equation of state  are similar to the phenomenological ones, evaluated in \cite{orl}. In particular, it seems that the introduction of the entanglement process, evaluated as a result of decoherence in the observable
universe, reproduces phenomenological equations of state of the universe, without assuming  any cosmological constant \emph{a priori}.

\section{conclusions and perspectives}

 Quantum decoherence can play an interesting role  in the context of observational cosmology. In particular, by assuming a decoherence signature at any epoch, it is possible to trace back the evolution of dark energy by comparing the variation of the Von Neumann entropy with the thermodynamic entropy. This approach can be pursued for any perfect fluid consistent with a given equation of state. Here we have considered  a FRW universe. In particular, we discussed   cosmological models where dark energy behavior can be achieved  without postulating \emph{a priori} a cosmological constant. The existence of dark energy at our epoch can  be seen as a quantum signature resulting  from a decoherence process, while along the universe evolution, it is possible to feature  different scaling volumes comparable with observational data. In particular, we investigated two cases: a volume $V\propto a^3$ and  a volume  $V\propto H^{-3}$. The corresponding cosmological models fairly well approximate the cosmological behavior at our time, showing that the existence of a cosmological constant appears as the limiting case of a more complicate dynamics.
Entanglement  between different cosmological eras leads to conclude that the existence of dark energy can be seen as the observable  effect of such process. In future researches, we are planning to develop in detail such an approach comparing observable cosmographic parameters with  thermodynamical  quantities related to information theory.

\end{document}